\documentclass[useAMS,usenatbib]{mnras}
\usepackage{times}
\usepackage{graphicx}

\usepackage{amsmath}
\usepackage{amssymb}

\usepackage[T1]{fontenc}
\usepackage{aecompl}
\usepackage{epstopdf}

\renewcommand{\vec}{\mathbf}

\begin{document}

\title[Self-consistent models of the first stars]{Self-consistent semi-analytic models of the first stars}

\author[E. Visbal et al.]{Eli Visbal$^{1}$\thanks{evisbal@flatironinstitute.org}, Zolt\'{a}n Haiman$^{2, 3}$, Greg L. Bryan$^{1,2}$ \\ $^1$Center for Computational Astrophysics, Flatiron Institute, 162 5th Ave, New York, NY, 10003, U.S.A. \\ 
$^2$Department of Astronomy, Columbia University, 550 West 120th Street, New York, NY, 10027, U.S.A. \\
$^3$Center for Cosmology and Particle Physics, New York University, 4 Washington Place, New York, NY, 10003, U.S.A. }

\maketitle

\begin{abstract}
We have developed a semi-analytic framework to model the large-scale evolution of the first Population III (Pop III) stars and the transition to metal-enriched star formation. Our model follows dark matter halos from cosmological N-body simulations, utilizing their individual merger histories and three-dimensional positions, and applies physically motivated prescriptions for star formation and feedback from Lyman-Werner (LW) radiation, hydrogen ionizing radiation, and external metal enrichment due to supernovae winds. This method is intended to complement analytic studies, which do not include clustering or individual merger histories, and hydrodynamical cosmological simulations, which include detailed physics, but are computationally expensive and have limited dynamic range. Utilizing this technique, we compute the cumulative Pop III and metal-enriched star formation rate density (SFRD) as a function of redshift at $z \geq 20$. We find that varying the model parameters leads to significant qualitative changes in the global star formation history. The Pop III star formation efficiency and the delay time between Pop III and subsequent metal-enriched star formation are found to have the largest impact. The effect of clustering (i.e. including the three-dimensional positions of individual halos) on various feedback mechanisms is also investigated. The impact of clustering on LW and ionization feedback is found to be relatively mild in our fiducial model, but can be larger if external metal enrichment can promote metal-enriched star formation over large distances.
\end{abstract}


\section{Introduction}
The standard model of cosmology has enabled theoretical predictions for the formation of the first stars in the Universe \citep[for a recent review see][]{2015ComAC...2....3G}.  These first Pop III stars are expected to form in pristine ${\sim}10^{5-6}~M_\odot$ dark matter halos known as ``minihalos'' at very high redshift. Because essentially no metals are present, gas in minihalos can only cool through molecular hydrogen transitions, and cannot achieve the low temperatures possible with metal cooling. This results in a higher Jeans mass in Pop III star forming gas compared to gas forming metal-enriched stars. Therefore, the first stars are thought to be more massive than metal enriched stars formed later in the Universe, with typical masses of $\sim 50-1000~M_\odot$ predicted by simulations \citep[e.g.][]{2015MNRAS.448..568H}.

In addition to studying the properties of individual Pop III stars, it is interesting to predict the total abundance of Pop III star formation as a function of cosmic time and the transition to metal enriched star formation. Ultraviolet (UV) and X-ray photons produced by Pop III stars and their remnants can potentially have an important impact on the thermal and ionization history of the intergalactic medium (IGM) in the early Universe. Thus, the abundance of Pop III stars has implications for high-redshift (e.g. $z\sim20$) radio observations of the redshifted 21cm line of neutral hydrogen \citep[e.g.][]{2012Natur.487...70V, 2013MNRAS.432.2909F} and measurements of the optical depth and polarization of the cosmic microwave background (CMB) \citep[e.g.][]{2003ApJ...583...24K, 2003ApJ...595....1H,  2015MNRAS.453.4456V, 2017MNRAS.467.4050M}. Additionally, if Pop III stars form in large numbers, they may explain the black hole binaries detected by the \emph{Laser Interferometer Gravitational-Wave Observatory (LIGO)} \citep{2016PhRvX...6d1015A}, and also produce a corresponding stochastic GW background detectable with the advanced \emph{LIGO+Virgo} network \citep{2016MNRAS.461.2722I}. Finally, the abundance of Pop III stars is important for ``stellar archaeology'' of very low metallicity stars in the Milky Way which could have formed early in the Universe \citep[e.g.][]{2015MNRAS.447.3892H, 2016MNRAS.460L..74H}.

Computing the global evolution of Pop III star formation is complicated by several different feedback processes. The most important of these processes is feedback from Lyman-Werner (LW) radiation produced by stars, which is able to photo-dissociate the molecular hydrogen needed for cooling in minihalos \citep[][]{1997ApJ...476..458H, 2000ApJ...533..594C, 2001ApJ...548..509M,2007ApJ...671.1559W,2008ApJ...673...14O,2011MNRAS.418..838W,2014MNRAS.445..107V}. Hydrodynamic cosmological simulations have found that LW feedback effectively increases the minimum mass of minihalos which can host Pop III star formation, with higher LW flux corresponding to higher minihalo mass required for star formation. Previous generations of Pop III stars produce a background of LW radiation which prevents subsequent star formation in low mass minihalos. Eventually the LW background is high enough to prevent star formation in all minihalos and star formation can only proceed in halos with virial temperatures greater than $\approx 10^4~{\rm K}$, where atomic hydrogen cooling can operate efficiently. Note that if Pop III star formation is dominated by stars that are only a few $M_\odot$, infrared photons may be more important for feedback than LW photons \citep{2012MNRAS.425L..51W}, however we assume that Pop III stars are massive throughout this paper.

 Ionizing radiation can also impact Pop III star formation by photoevaporating gas out of minihalos \citep{2004MNRAS.348..753S, 2005MNRAS.361..405I} and suppressing metal-enriched star formation in larger dark matter halos \citep[e.g.][]{1994ApJ...427...25S,1996ApJ...465..608T,1998MNRAS.296...44G,2000ApJ...542..535G,2004ApJ...601..666D,2006MNRAS.371..401H,2008MNRAS.390..920O,2013MNRAS.432L..51S,2014MNRAS.444..503N, 2017MNRAS.469.1456V}. Additionally, metals produced in stars and spread throughout the Universe via supernovae winds can pollute minihalos, preventing Pop III star formation and triggering metal-enriched star formation \citep{2001MNRAS.328..969B, 2005ApJ...626..627O, 2012ApJ...745...50W, 2015MNRAS.452.2822S}. 

Previous efforts to predict the global evolution of the first stars have generally taken one of two approaches, analytic modeling usually based on the extended Press-Schechter formalism \citep[e.g][]{2005ApJ...623....1M, 2007ApJ...659..890W, 2009ApJ...694..879T,2006ApJ...650....7H, 2015MNRAS.453.4456V, 2016MNRAS.462.3591M} or hydrodynamic cosmological simulations \citep[e.g.][]{2010MNRAS.407.1003M, 2015ApJ...807L..12O, 2016ApJ...823..140X}, each of which has its benefits and drawbacks. Generally speaking, analytic work can test many model parameterizations very quickly, but information about individual halo merger histories and spatial clustering of halos is not included (though see \cite{2003ApJ...589...35S} for an analytic treatment which does include clustering). On the other hand, hydrodynamical simulations can include detailed physics, but are computationally expensive, permitting only relatively small boxes compared to the horizon of LW photons and sparse sampling of the uncertain model parameter space. Additionally, N-body simulations with radiative transfer have simulated reionization with Pop III stars by including minihalos as a subgrid prescription \citep{2012ApJ...756L..16A}, but this cannot carefully account for local feedback processes on small scales. Another approach has been to perform large-scale semi-numerical modeling, where individual halos are not considered, but fluctuations in the LW background are included on scales encompassing the LW photon horizon \citep{2012Natur.487...70V, 2013MNRAS.432.2909F}.

In this paper we develop a new semi-analytic technique to study the evolution of the first Pop III stars and the transition to metal-enriched star formation. We utilize dark matter-only N-body simulations (which resolve individual minihalos) and then apply a prescription to include Pop III and metal-enriched star formation. This is similar in some respects to semi-analytic models of galaxy formation used at lower redshifts \citep{2014ApJ...795..123L}.  The prescription presented here includes LW and ionization feedback as well as external metal enrichment (i.e. metals coming from a halo which is not a progenitor of the halo being enriched) due to supernovae winds. These semi-analytic simulations contain both merger history and clustering information not present in analytic approaches, but are much less computationally expensive than full hydrodynamical simulations. This allows us to survey a wide range of physics at a relatively low computational cost. We utilize our model to test a variety of different parameterizations of early star formation and feedback mechanisms. 

We find that varying the assumptions regarding the physics of the first stars, in particular the Pop III star formation efficiency and the delay time between Pop III star formation and subsequent metal-enriched star formation in the same halo, can result in qualitative changes to the large-scale evolution of the SFRD. Additionally, we use our model to test the impact of the small-scale clustering of individual halos on feedback effects, which has been suggested as potentially important \citep{2006ApJ...649..570K}.

We note that our approach is similar to those of \cite{2009ApJ...700.1672T},  \cite{2012MNRAS.425.2854A}, and \cite{2013ApJ...773..108C}, however here we consider a more detailed model, which extends these studies by simultaneously including spatial fluctuations in LW feedback, ionization feedback, and external metal enrichment.

This paper is structured as follows. In \S2 and \S3, we describe the N-body simulations utilized in the model and the details of our semi-analytic model, respectively. We present the results for a range of parameters and discuss the importance of clustering in \S4 and discuss comparisons with previous work in \S5. Finally, we summarize and discuss our conclusions in \S6. Throughout, we assume a $\Lambda {\rm CDM}$ cosmology with parameters consistent with \cite{2014A&A...571A..16P}: $\Omega_{\rm m} = 0.32$, $\Omega_{\Lambda} = 0.68$, $\Omega_{\rm b} = 0.049$, $h=0.67$, $\sigma_8=0.83$, and $n_{\rm s} = 0.96$.

\section{N-body simulations}
\begin{figure}
\includegraphics[width=88mm]{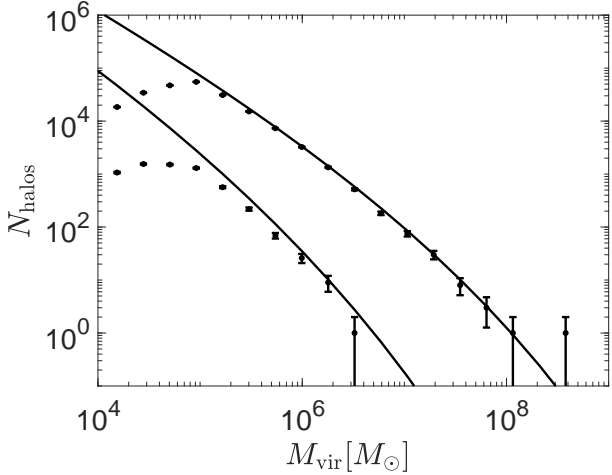}
\caption{\label{mass_func} The number of halos found in our box (black points with error bars) as a function of halo virial mass and the number expected from the Sheth-Tormen mass function (black curves). The mass bins are width $\Delta M_{\rm vir} = 0.58M_{\rm vir}$ and Poisson error bars are shown ($\sigma = \sqrt{N_{\rm halo}}$). The upper curve and points are for $z=20$ and the lower are for $z=30$.    } 
\end{figure}

We produce the cosmological N-body simulations needed for our model with the publicly available code \textsc{gadget2} \citep{2001NewA....6...79S}. For the results discussed below, our simulations contained $1024^3$ particles in a $5~{\rm Mpc}$ box. This corresponds to a particle mass of $4600~M_\odot$. Thus, $10^5~M_\odot$ minihalos are resolved with $\sim 20$ particles each.  A gravitational softening length of 160 pc was used. Initial conditions were generated with \textsc{MUSIC} \citep{2011MNRAS.415.2101H} using second order Lagrangian perturbation theory.
 The simulation was started at $z=200$ and we take snapshots at $\Delta z = 1$ from $z=40-20$. The upper limit of this redshift range corresponds to the point when the first stars in our simulation box form (a larger box would have the first star at higher redshift). By the lower limit of this range, metal-enriched star formation dominates over Pop III in the model parameterizations we have explored. To build merger trees, we use \textsc{rockstar} \citep{2013ApJ...762..109B} and \textsc{consistent trees} \citep{2013ApJ...763...18B}. In Figure \ref{mass_func}, we show the number of halos found in our simulation compared to estimates from the Sheth-Tormen  mass function \citep{1999MNRAS.308..119S}. At $z=20$ we find very close agreement with our simulations and Sheth-Tormen, but at higher redshift we find that this mass function overpredicts the number of halos by a relatively modest amount (e.g. a factor of ${\sim}2$ at $z=30$). Throughout, we use the virial mass definition from \cite{1998ApJ...495...80B}. 

\section{Semi-analytic model}
\subsection{Basic framework}
\begin{figure*}
\includegraphics[width=88mm]{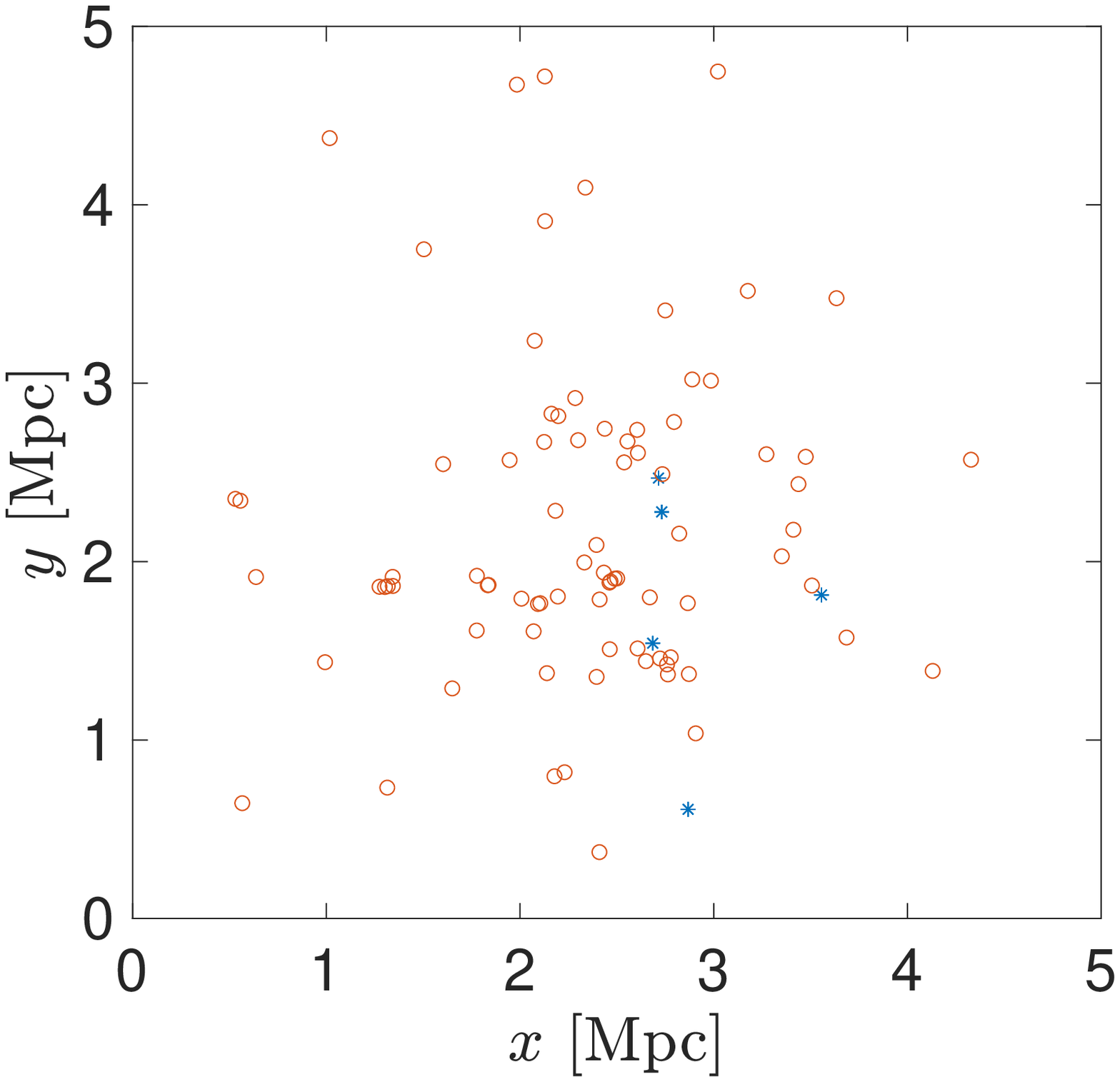}
\includegraphics[width=88mm]{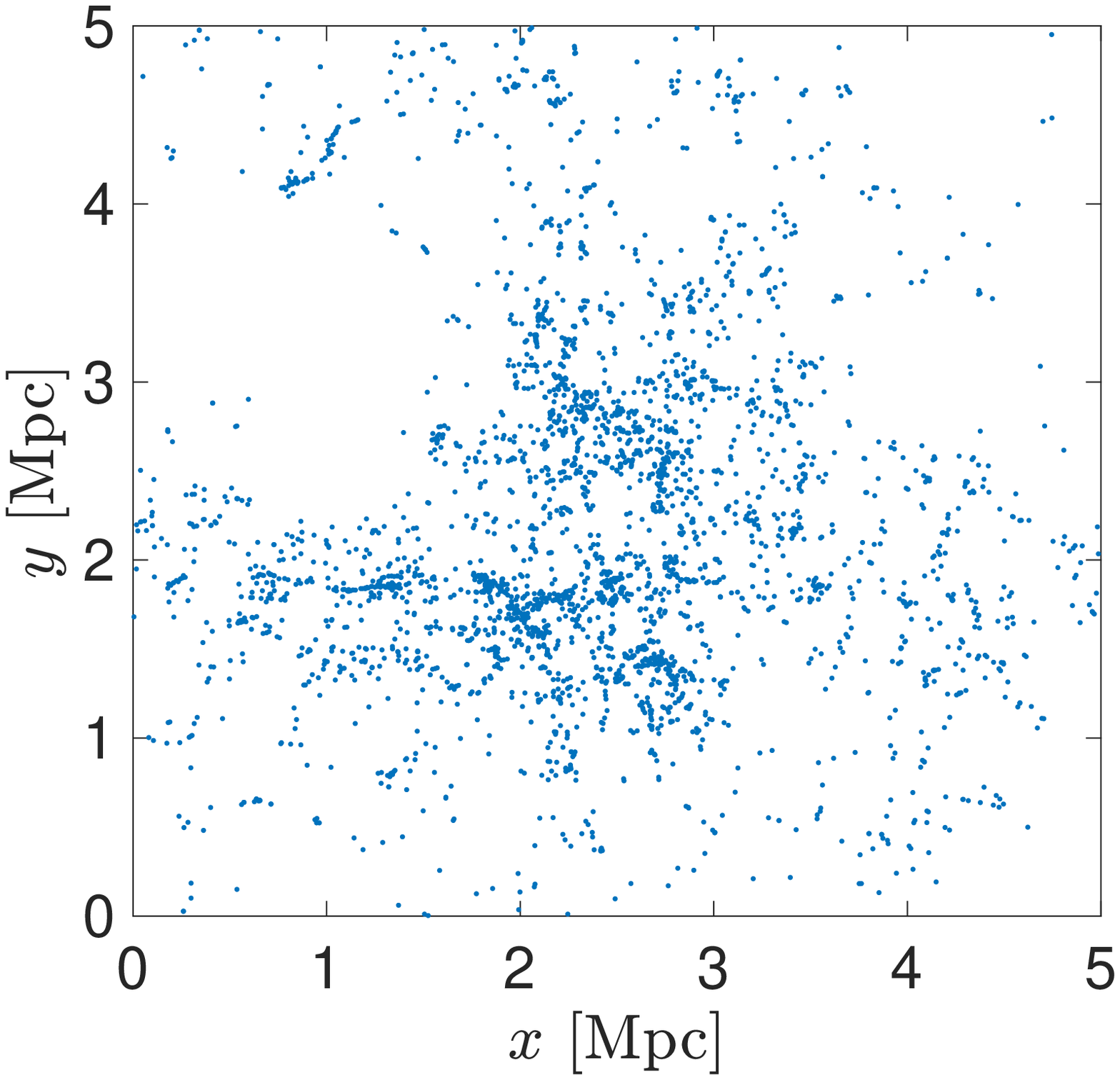}
\caption{\label{slice}
Projections of dark matter halos in our semi-analytic model with the fiducial parameters described in \S4. The left panel shows the location of metal-enriched (circles) and Pop III (*'s) star-forming halos at $z=20$, while the right panel shows the location of all halos above $10^6~M_\odot$. Halos with metal-enriched star formation had progenitors with Pop III stars in the past.  } 
\end{figure*}

We have developed a semi-analytic model to study the large-scale properties of the first Pop III stars and the transition to metal-enriched star formation. Our model utilizes merger trees from N-body simulations including the three-dimensional positions of dark matter halos. Prescriptions for star formation and feedback are applied to these merger trees to determine the star formation history throughout the simulation box as a function of redshift. In Figure \ref{slice}, we show a projection of the metal-enriched and Pop III star-forming halos in this model for the fiducial case described below.

In order to model star formation and feedback at high temporal resolution, we use a simple scheme to track dark matter halos between output snapshots of the N-body simulation. We divide the period between consecutive snapshots into many equally spaced time steps. This is necessary to accurately track the changes in LW flux due to Pop III stars turning on and off. The lifetime of a massive Pop III star is a few Myr, while towards the end of our simulations the difference between simulation snapshots is over 10 Myr. We have adopted time steps between snapshots of $\sim 0.25$ Myr at $z \sim 20$ (and smaller at higher redshift), which allows us to resolve changes in LW flux on timescales significantly shorter than the Pop III lifetime.

For a halo which does not experience any mergers between snapshots, the mass of the halo grows (or shrinks in some cases due to interactions with nearby halos) linearly with time such that it has the correct mass at the later snapshot. For cases with mergers (i.e. more than one progenitor halo ending in a single descendant halo), we draw random numbers, sampled evenly in the time between simulation snapshots, to determine when mergers occur. For a set of $N_{\rm p}$ common progenitors, we generate $N_{\rm p} -1$ random times. The first halo in the set merges into the second at the first time, the second halo (now containing the first) merges with the third at the second time, and so on. Whenever two progenitor halos merge, the mass growth rate of the merged pair becomes the sum of the progenitors' growth rates. 

We also assign velocities to halos and track changes in their positions between simulation snapshots. For halos without mergers, we simply assign a constant velocity chosen to transport the halo from its position in one snapshot to its location in the next. For halos with mergers, we assign constant velocities to each progenitor such that each halo meets the halo it merges with at the time of the (randomly assigned) merger. For a set of merging progenitors, when the first halo merges into the second, the merged pair keeps the velocity of the second, when the second merges with the third, the merged pair takes the velocity of the third, and so forth for the rest of the set. Note that consistent with our N-body simulations described below, we consider periodic boundary positions (i.e. a halo that reaches one edge of the box will re-emerge on the other side). Throughout our model, we do not track the substructure of halos. If a subhalo merges with a halo and then subsequently leaves the halo in a later snapshot, we assume that subhalo (distinct halo once it leaves) does not produce any stars. We do this because the subhalo may have had gas stripped during the merger and because we have already assigned it some star formation when it initially merged. 

We emphasize that the sub-snapshot prescription described here is not intended to be strictly physical, rather it is intended to allow us to run at very high time resolution without generating large numbers of N-body simulation outputs. We find that when changing the random seeds of the mergers, the large-scale features of our model (e.g. metal-enriched and Pop III star formation rate density as a function of redshift) remain mostly unchanged. We also find that turning the sub-snapshot velocities off (leading to instantaneous jumps in positions when mergers occur) leaves our results essentially unchanged. This suggests that the somewhat arbitrary details of the sub-snapshot physics we have chosen do not have a strong impact on global properties of star formation. Now, we will describe the prescriptions for star formation and the various feedback processes which are included in our model. 
 
\subsection{LW feedback}

Following the results of cosmological simulations \citep{2001ApJ...548..509M, 2007ApJ...671.1559W, 2008ApJ...673...14O}, we assume that Pop III star formation occurs in pristine dark matter halos once they reach a critical mass which depends on the redshift and local LW flux. This mass scale is given by
\begin{equation}
M_{\rm crit} = 2.5 \times 10^5 \left (\frac{1+z}{26} \right )^{-3/2} \left ( 1 + 6.96  \left [4\pi J_{\rm LW, 21} \right ]^{0.47} \right ) M_\odot,
\end{equation}
where $J_{\rm LW, 21}$ is LW intensity in units of $10^{-21} {\rm erg~s^{-1}~cm^{-2}~Hz^{-1}~sr^{-1}}$ \citep{2013MNRAS.432.2909F}. If $M_{\rm crit}$ exceeds the atomic cooling threshold, given by 
\begin{equation}
M_{\rm a} = 5.4\times 10^7 \left ( \frac{1+z}{11} \right )^{-3/2}~M_\odot,
\end{equation}
$M_{\rm a}$ is used for the critical mass instead. This choice of $M_{\rm a}$ corresponds to the typical minimum
mass of haloes that are able to cool in the absence of molecular hydrogen in the
simulations of \cite{2014MNRAS.439.3798F}. 

For each pristine halo at each redshift step, we compute the LW intensity as $J_{\rm LW, 21} = J_{\rm loc}(\vec{x}, z) + J_{\rm bg}(z)$, where $J_{\rm loc}(\vec{x}, z)$ is the local flux from nearby individual sources and $J_{\rm bg}(z)$ is the mean LW background on large scales. The local flux is given by
\begin{equation}
J_{\rm loc}(\vec{x}, z) = \sum_{i} \frac{L_{\rm LW, i}}{(4 \pi)^2 \Delta \nu_{\rm LW} | \vec{x} - \vec{x}_i |^2}, 
\end{equation} 
where the summation is over all star-containing halos within comoving distance $R_{\rm max} = \sqrt{2}/2 L_{\rm box}$ and $L_{\rm box}$ is the N-body simulation box size. $L_{\rm LW, i}$ is the LW luminosity of the $i$'th halo, $\Delta \nu_{\rm LW} = 5.8 \times 10^{14}{\rm Hz}$ is the frequency range of LW photons (corresponding to the difference between 11.2 and 13.6 eV photons), and $\vec{x}_i$ is the position of the $i$'th halo (in physical units). To ensure sources at the edge of the box do not have artificially low values of $J_{\rm loc}$, we use periodic boundary conditions when computing this quantity (this is also why we have all halos see sources out to the same distance, $R_{\rm max}$). Note that for each halo, its observed $J_{\rm loc}$ is the flux from nearby halos closer than $R_{\rm max}$ and its observed $J_{\rm bg}$ is the flux from halos farther then $R_{\rm max}$ (making the simplifying assumption that this contribution is spatially uniform with a redshift-dependent intensity taken from the evolving mean emissivity of the box).

The uniform background component of the LW flux is given by 
\begin{equation}
\label{bg_eqn}
J_{\rm bg}(z) = \frac{c (1+z)^3}{4 \pi} \int_{z_{\rm Rmax}}^\infty dz' \epsilon_{\rm LW}(z') \left | \frac{dt_{\rm H}}{dz'} \right | f_{\rm LW}(z', z),
\end{equation}
where $\epsilon_{\rm LW}(z')$ is the mean LW emissivity in our box as a function of redshift, $t_{\rm H}$ is the Hubble time, and $f_{\rm LW}(z',z)$ the attenuation of LW flux observed at redshift $z$ from sources at redshift $z'$ due to LW photons being redshifted into Lyman series resonance lines and absorbed. We approximate this attenuation with Eq. 22 in \cite{2009ApJ...695.1430A}. The mean emissivity is given by
\begin{equation}
\epsilon_{\rm LW}(z)=\frac{E_{\rm LW}}{\Delta \nu_{\rm LW} m_{\rm p}} \left [ \eta_{\rm III}{\rm SFRD}_{\rm III}(z) + \eta_{\rm II}{\rm SFRD}_{\rm II}(z) \right ],
\end{equation}
where $E_{\rm LW}= 2 \times 10^{-11}~{\rm erg}$ is the energy of a LW photon, $m_{\rm p}$ is the proton mass, ${\rm SFRD}_{\rm III}$ and ${\rm SFRD}_{\rm II} $ are the Pop III and metal-enriched star formation rate densities, and $\eta_{\rm III}$ and $\eta_{\rm II}$ are the number of LW photons produced per baryon incorporated in Pop III and metal-enriched stars (discussed below).

The limit of integration $z_{\rm Rmax}$ corresponds to the redshift an observer at $z$ sees a source at $R_{\rm max} = \sqrt{2}/2 L_{\rm box}$ due to the finite light travel time. Note that when we compute the star formation for a particular time step, we use the sources from the previous time steps to compute the LW feedback. Our results converge as we make time steps small. 

For the highest redshifts in our simulations, Eq. \ref{bg_eqn} is not a good estimate for the LW background due to the low number of sources in the box and rapid fluctuations as these sources turn on and off. To address this, above $z_{\rm hi-z} = 30$ we assume the LW background evolves exponentially as $J_{\rm bg}(z) = A \exp(-B z)$. The parameters $A$ and $B$ are fit iteratively by computing $J_{\rm bg}(z)$ with Eq. \ref{bg_eqn} from the previous iteration's fit and then re-fitting $A$ and $B$ to the new background. This fitting process is continued until changes are small from iteration to iteration. This iteration is required because one does not know \emph{a priori} the best values of $A$ and $B$ to assume, thus for the first few iterations the assumed $J_{\rm bg}(z)$ can be much different than the value one gets from fitting the background computed with Eq. \ref{bg_eqn}.

\subsection{Star formation}
The first time a pristine halo exceeds $\min(M_{\rm crit}, M_{\rm a})$ we introduce Pop III stars. For simplicity, we assume that all Pop III stars are of the same mass, $M_{\rm PopIII}$. Our method can easily be modified in future work to incorporate a distribution of stellar masses sampled from an initial mass function (IMF). The total mass of Pop III stars assigned to a halo is 
\begin{equation}
M_{\rm III, *} = \max(M_{\rm PopIII}, f_{\rm III}\frac{\Omega_{\rm b}}{\Omega_{\rm m}}M_{\rm h}), 
\end{equation}
where $f_{\rm III}$ is the Pop III star formation efficiency and $M_{\rm h}$ is the total dark matter halo mass (note that $f_{\rm III} = 0$ implies one Pop III star per halo). If $f_{\rm III}>0$, massive pristine halos can produce multiple stars. In this case, halos can have a total stellar mass which is equivalent to a fractional number of stars. This choice was made for simplicity and in these cases the number of ionizing/LW photons produced is proportional to the stellar mass formed in Pop III stars. In future work we intend to adopt a more sophisticated treatment of the IMF.

We assume the stellar luminosity is constant over the lifetime of the Pop III stars, $t_{\rm PopIII}$, and characterize it in terms of $\eta_{\rm III}$, the number of LW and ionizing photons (assumed to be the same) per baryon forming stars. Both $t_{\rm PopIII}$ and $\eta_{\rm III}$ depend on the value of $M_{\rm PopIII}$ and are estimated from the calculations of \cite{2002A&A...382...28S}.

After the death of a Pop III star and possible corresponding supernova, we assume there is some recovery time before the gas can fall back into a halo and begin metal-enriched star formation. Metal-enriched star formation is included in a halo $t_{\rm delay}$ after the earliest Pop III star(s) formed in all of its progenitor halos. Once metal-enriched star formation is present, we assume a star formation rate (SFR) proportional to the mass growth rate of the halo with an additional boost produced by the merger of pristine halos. Thus, the metal-enriched SFR at a particular time step is given by
\begin{equation}
\label{popII_eqn}
{\rm SFR_{\rm II}} = f_{\rm II} \frac{\Omega_{\rm b}}{\Omega_{\rm m}} \dot{M}_{\rm h} + f_{\rm II} \frac{\Omega_{\rm b}}{\Omega_{\rm m}} \frac{{M}_{\rm merge}}{\Delta t},
\end{equation}
where $f_{\rm II}$ is the metal-enriched star formation efficiency, $\dot{M}_{\rm h}$ is the growth of the dark matter halo due to unresolved accretion discussed above, ${M}_{\rm merge}$ is the total mass of pristine halos which merged with the halo at the time step, and $\Delta t$ is the length of the time step.

To compute the LW and ionizing flux of metal-enriched stars we assume that, at a given time, all stars which formed within $t_{\rm PopII}$ emit at a constant luminosity. Thus, the luminosity is given by
\begin{equation}
L_{\rm II}(z)  =  \int_{t_{\rm H}(z) - t_{\rm PopII}}^{t_{\rm H}(z)}  dt'  \frac{{\rm SFR_{II}}(t') \eta_{\rm II}}{t_{\rm PopII} m_{\rm p}},
\end{equation}
where $\eta_{\rm II}$ is the number of ionizing/LW photons produced per baryon in metal-enriched stars and $m_{\rm p}$ is the mass of the proton. The parameter $t_{\rm PopII}$ is meant to represent the lifetime of massive stars which produce most of the ionizing/LW radiation.

\subsection{Ionization feedback}
In regions where the IGM has been ionized, we expect the minimum mass of halos hosting star formation to be significantly higher than in neutral regions. We use a simple prescription similar to the approach of \cite{2004ApJ...613....1F} to treat this ionization feedback. To determine if a halo is in an ionized region, we examine a series of comoving spheres of around the halo. For each sphere, we identify all halos contained in the sphere that are producing or whose progenitors produced ionizing radiation and compute the size of the HII region these sources would create. If at the current redshift, the HII region is larger than the sphere (with radius $R_{\rm s}$) we assume the halo is in an ionized region. The size of the HII region, $R_{\rm i}$, is computed with
\begin{equation}
\label{reion_eq}
\frac{dR^3_{\rm i}}{dt} = 3H(z)R^3_{\rm i} + \frac{3\dot{N}_\gamma}{4\pi \langle n_H \rangle } - C(z) \langle n_H \rangle \alpha_{\rm B} R^3_{\rm i},
\end{equation}
where $\dot{N}_\gamma$ is the rate of ionizing photon production in the sphere as a function of time, $H(z)$ is the Hubble parameter, $\alpha_{\rm B}= 2.6\times 10^{-13} {\rm cm^3s^{-1}}$ is the case~B recombination coefficient of hydrogen at $T=10^4$ K, $\langle n_H \rangle$ is the mean cosmic hydrogen density, and $C(z) \equiv \langle n_{\rm HII}^2 \rangle / \langle n_{\rm HII} \rangle^2 $ is the clumping factor of the ionized IGM. We assume a clumping factor of 
\begin{equation}
C(z) = 2 \left ( \frac{1+z}{7}  \right )^{-2} + 1.
\end{equation} 
This formula gives values similar to that found in the Illustris
simulation for gas below 20 times the mean baryon density \citep{2015MNRAS.453.3593B}. It is also similar to the values found in \citep{2012MNRAS.427.2464F}.

We start with a sphere of radius $R_{\rm s}$ equal to the halo virial radius and check spheres from this radius up to the entire size of the box. We increase the number of spheres used until we achieve convergence. To compute the amount of photons, we assume the same quantity as for the LW photons described above, but reduced by escape fractions of $f_{\rm esc, II}$ and $f_{\rm esc, III}$ for halos with metal-enriched and Pop III star formation, respectively.

If a pristine halo is in an ionized bubble, the minimum mass of star formation is increased to $M_{\rm ion}$, which we assign a fiducial value of $M_{\rm ion} = 1.5 \times 10^8 \left ( \frac{1+z}{11}  \right )^{-3/2}~M_\odot$ \citep[consistent with the results of][]{2004ApJ...601..666D}. For halos which are already undergoing metal-enriched star formation, we assume that if they reside in an ionized bubble, gas can no longer fall in and star formation is quenched on the dynamical timescale of the halo. In this case, the SFR is attenuated as
\begin{equation}
{\rm SFR_{\rm bub}} = {\rm SFR_{\rm II}} \exp \left (  - \Delta t_{\rm bub}/t_{\rm dyn}     \right ),
\end{equation}
where $\Delta t_{\rm bub}$ is the time the halo has been in an ionized region and $t_{\rm dyn}$ is the dynamical time at the mean density of the halo, which we approximate as  $t_{\rm dyn} \approx 0.1t_{\rm H}$. For these halos, we also assume that there is no pristine merger component to the metal-enriched SFR due to photoevaporation (i.e. we set the second term in Eq. \ref{popII_eqn} to zero).

\subsection{Metal enrichment}
We also include external metal enrichment in our semi-analytic model by considering the size of metal enriched bubbles produced by supernovae explosions. 
Following \cite{2009ApJ...700.1672T}, we assume that after the death of a star, a metal enriched bubble expands at velocity $v_{\rm bub} = f_{\rm bub} \times 60~{\rm km~s^{-1}}$ until it reaches a comoving radius of $R_{\rm bub} = f_{\rm bub} \times 150~h^{-1}~{\rm kpc}$. Here $f_{\rm bub}$ is a parameter that we switch between 1 and 0 to turn external enrichment on and off. We find that the bubble radius given by this procedure is in rough agreement with the simulations of \cite{2015MNRAS.452.2822S}. 

Pristine halos that fall within the radius of a metal bubble are assumed to be metal enriched. The metallicity of these externally enriched halos is computed by assuming that the metals produced in supernovae are spread uniformly throughout the bubbles. If a halo is hit by more than one bubble, the metallicity is set to the sum of the metallicity of the bubbles. We assume $10~M_\odot$ of heavy elements are produced per each $40~M_\odot$ (chosen as the fiducial Pop III stellar mass) of Pop III stars \citep{2006NuPhA.777..424N}  and $1~M_\odot$ of metals are produced for every $100~M_\odot$ of metal-enriched stars \citep[this is the approximate metal yield assuming that stars above $8~M_\odot$ lead to a supernova and $\sim 1~M_\odot$ of heavy metals are produced per supernova][]{2001PhR...349..125B}. If the external enrichment becomes higher than a critical metallicity, $Z_{\rm crit}$, we impose a critical mass, $M_{\rm min, met}$, for new (metal-enriched) star formation in these halos and assume Pop III stars can no longer form. For regions with metallicity below this critical value we still allow Pop III star formation. For the calculations presented below we assume a fiducial value of $M_{\rm min, met} = 2 \times 10^5 ~ M_\odot$. There have been a number of previous efforts to determine the critical metallicity at which cooling leads to significantly smaller stellar masses than in Pop III star formation \citep[e.g.][]{2003Natur.425..812B, 2005ApJ...626..627O, 2007ApJ...661L...5S, 2009ApJ...691..441S}. These studies tend to find a critical value around $Z_{\rm crit} \approx 10^{-3.5} Z_\odot$, which we use as a fiducial value.  

In our fiducial model, we do not include external metal enrichment. Because there are many more potential star forming halos in runs when it is included, we slightly simplify the reionization feedback prescription described above to accelerate the computation. Rather than solve Eq. \ref{reion_eq}, we simply count the total number of ionizing photons in spheres around a halo and compare it to the number of hydrogen atoms in the spheres. If there are more photons than atoms in any sphere, we consider the halo to be in an ionized region. This gives very similar results to using Eq. \ref{reion_eq}, which implies that the overall impact of recombinations is small. This is because the rate at which the SFRD increases is much faster than the recombination rate. For example, the recombination time at $z=25$ is $\sim 40 ~ {\rm Myr}$ and $40 ~ {\rm Myr}$ after $z=25$ ($z\approx20$), the SFRD density has increased by an order of magnitude. Thus, most of the ionized photons have been produced more recently than the recombination time scale, causing recombination to have a relatively small impact.

\subsection{Fiducial Parameters}
Here we discuss the parameter choices for our fiducial model, which are listed in Table \ref{table}. For metal-enriched star formation, we have adopted an efficiency of $f_{\rm II}=0.05$, which we take from \cite{2015MNRAS.453.4456V}. This value was determined using abundance matching and the observed UV luminosity function of galaxies at $z\approx 6$ from \cite{2015ApJ...803...34B}. We have chosen an escape fraction for metal-enriched galaxies of $f_{\rm esc,II}=0.1$, which is compatible with the simulations of \cite{2014MNRAS.442.2560W} for the low-mass halos we follow here, and a number of ionizing photons per baryon of $\eta_{\rm II}=4000$, which corresponds to a Salpeter IMF from 0.1 to 100 $M_\odot$ and metallicity $Z=0.0004$ \citep[see table 1 in][]{2007MNRAS.377..285S}. We note that these values are quite uncertain and could also evolve with redshift, which for simplicity we do not explore in this paper. We also note that our simple assumption of equal numbers of ionizing and LW photons likely underestimates the LW flux by a factor of a few from metal-enriched galaxies. Given the comparable uncertainties in the many of the important parameters (e.g.~the evolution of the metal-enriched star formation efficiency), we defer a more careful treatment for future work. 

For simplicity we have assumed a single Pop III stellar mass of $M_{\rm PopIII} = 40$. We note that there is large uncertainty in Pop III masses and in future work we intend to sample from a number of different IMFs \citep[e.g. that from][]{2015MNRAS.448..568H}. Given our choice of $M_{\rm PopIII}$, we take the values of $\eta_{\rm III}$ and $t_{\rm PopIII}$ from the stellar models of \cite{2002A&A...382...28S}. We assume that our Pop III stars end their lives as supernovae leading to $t_{\rm delay} = 10^7 ~{\rm yr}$ before the gas can settle back into the halo and form metal-enriched stars. This value is consistent with the simulations of \cite{2014MNRAS.444.3288J}, but we note that it can change as a function of supernovae energy and halo mass. Below, we explore increasing the value of this delay which could correspond to larger Pop III stars leading to more energetic supernovae. In future work, we will explore making the delay vary from halo to halo based on the Pop III masses, halo mass, and redshift.

In our fiducial model we assume there is no external metal enrichment, $f_{\rm bub}=0$. This may correspond to the case where much of the metals from winds do not efficiently mix with dense gas already inside minihalos \citep{2008ApJ...674..644C}. As noted above, $f_{\rm bub}=1$ gives metal bubbles with radii similar to those found in the simulations of  \cite{2015MNRAS.452.2822S}. We regard this as a realistic possibility and include it in our parameter variations presented below.

\section{Results}
In this section, we study how various physical processes in our semi-analytic model impact the large-scale properties of early star formation. 
We have first simulated a fiducial case, with parameter values listed in Table \ref{table}, as discussed in the previous section. 
We emphasize that there is great uncertainty in the most realistic values of these parameters, therefore the results here are not intended to be a precise description of the first stars, but instead are meant to show how changes in various parameters can qualitatively alter their global properties. We note that we have used 50 timesteps (evenly spaced in time) between each of our N-body simulation snapshots (taken with $\Delta z = 1$), and that higher resolution does not significantly change our results.

\subsection{Parameter variations}
\begin{figure*}
\includegraphics[width=88mm]{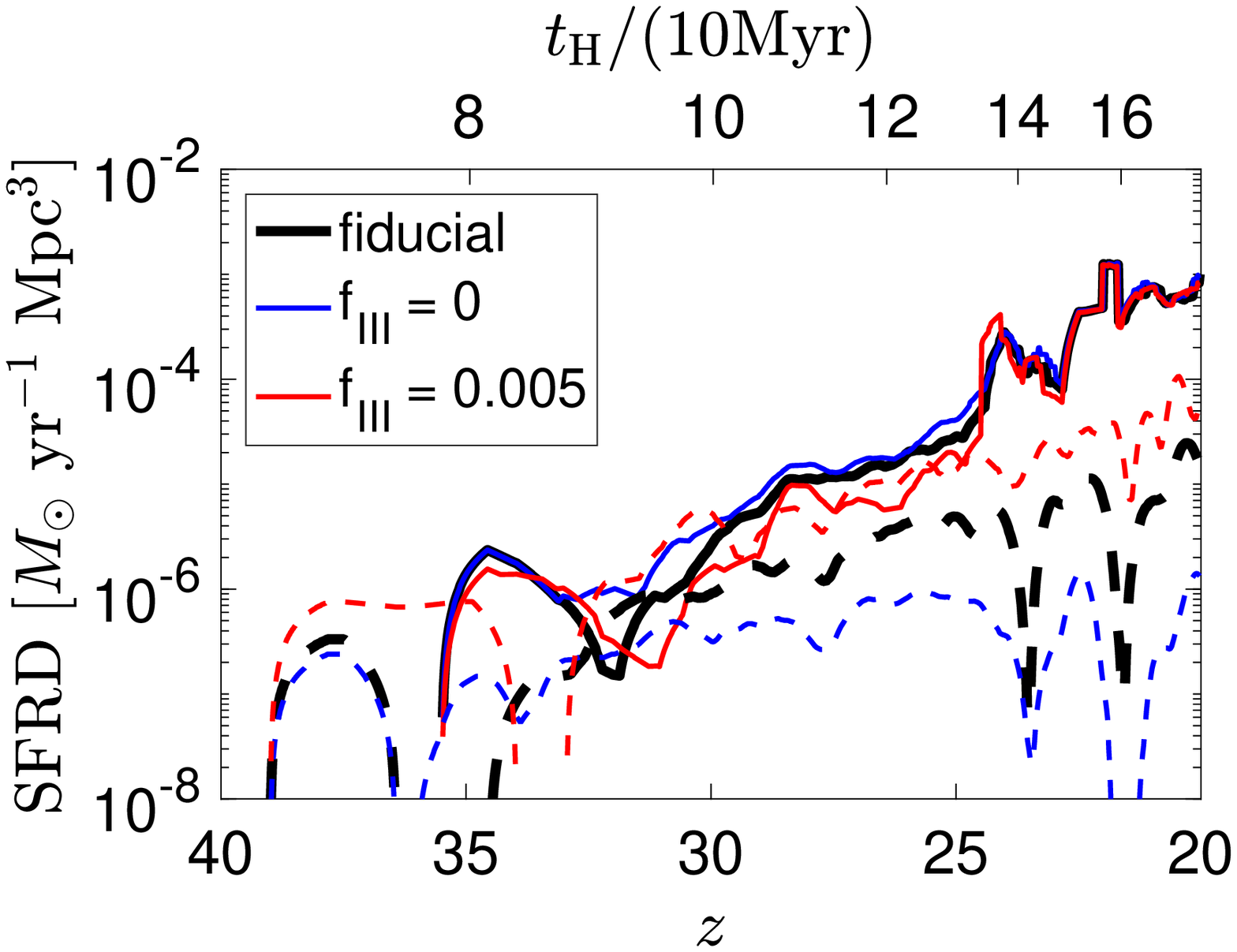}
\includegraphics[width=88mm]{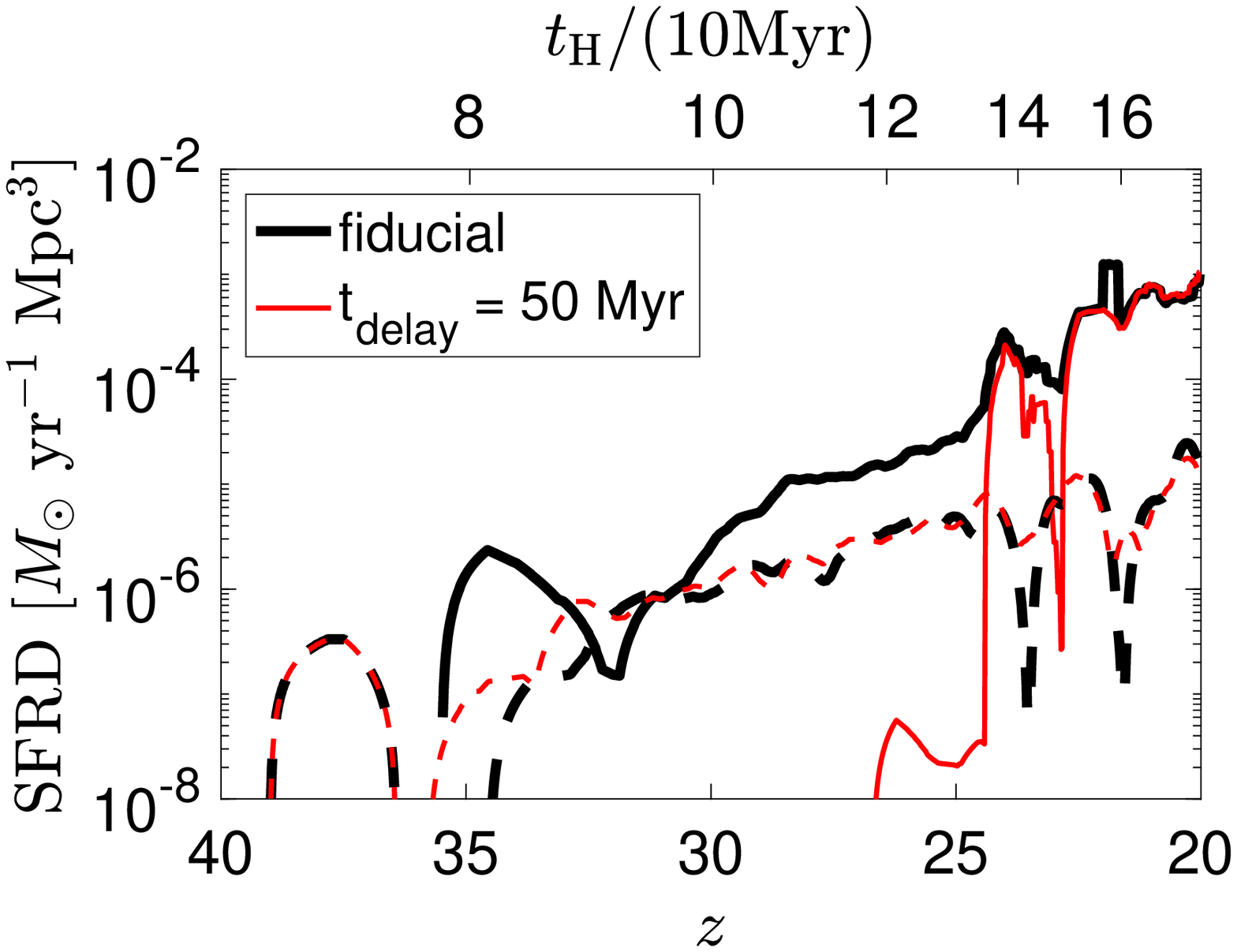}
\includegraphics[width=88mm]{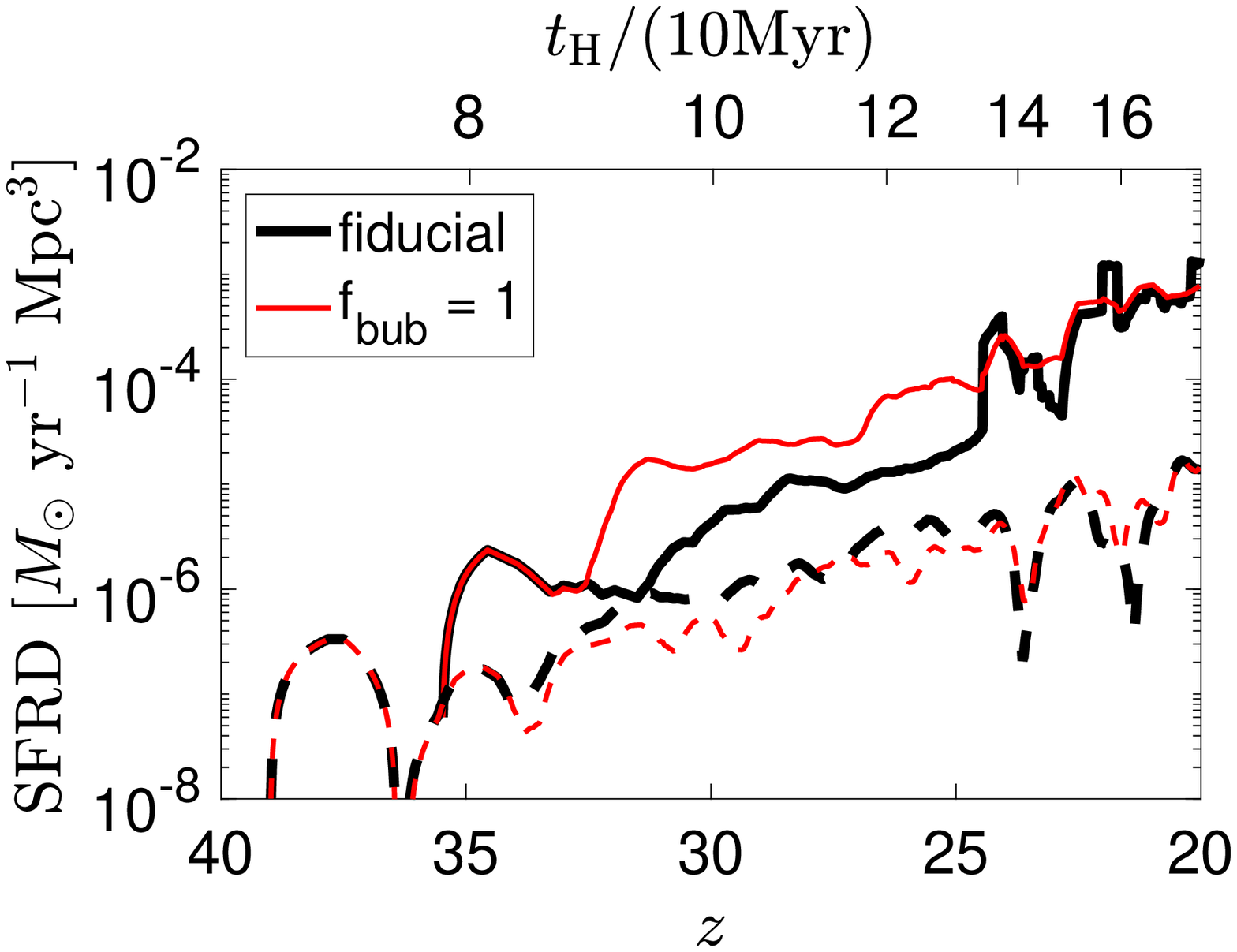}
\includegraphics[width=88mm]{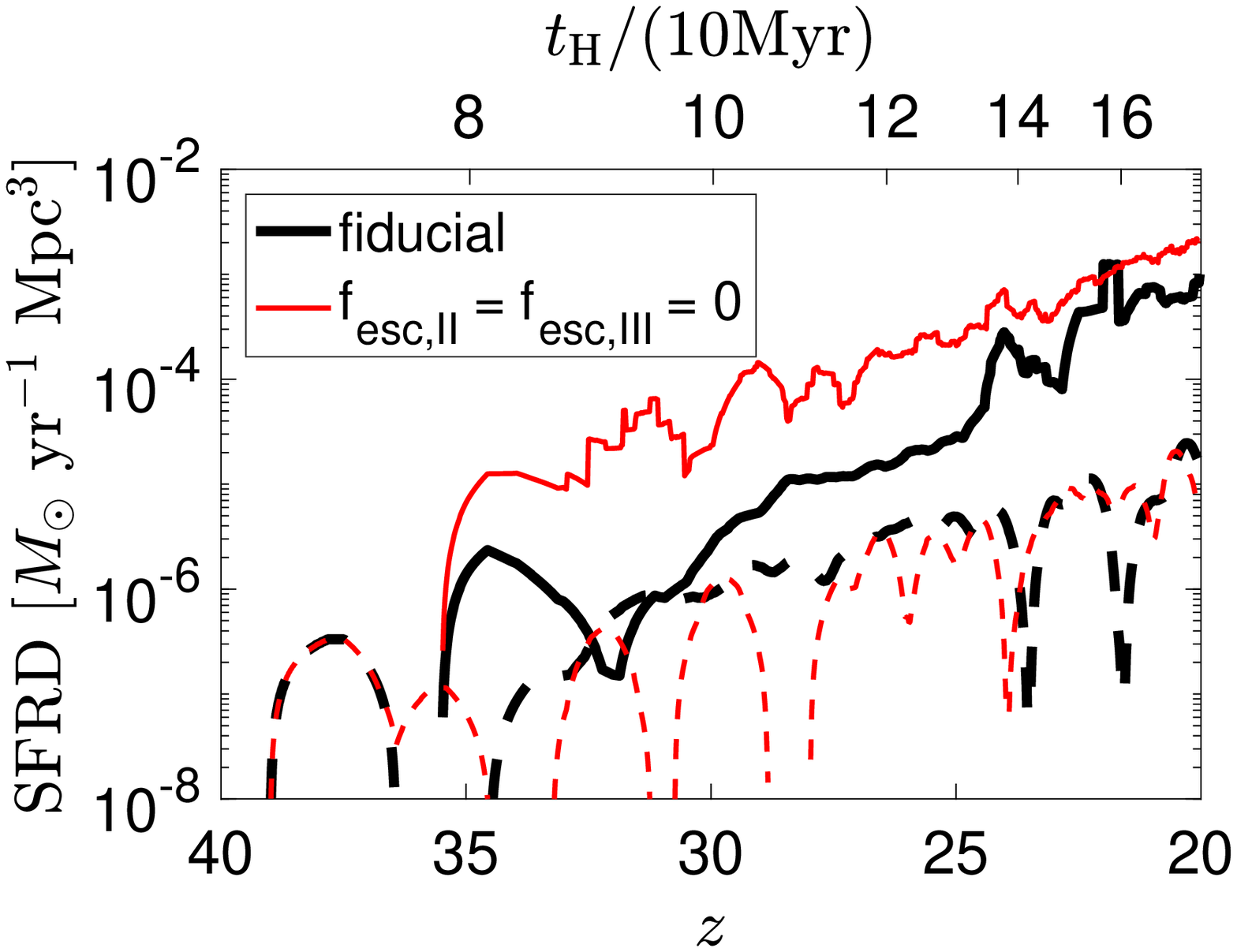}
\caption{\label{parameters}
Pop III (dashed curves) and metal-enriched (solid curves) SFRDs for variations in a number of model parameters. The SFRDs have been smoothed on a scale of $10^7$ yr.} 
\end{figure*}

In Figure \ref{parameters}, we plot the SFRD of Pop III and metal-enriched stars as a function of redshift, comparing our fiducial model to cases with changes in $f_{\rm III}$, $t_{\rm delay}$, $f_{\rm bub}$, and the ionizing escape fractions for both metal-enriched and Pop III star-forming halos. To make changes more clear we have smoothed the instantaneous SFRD, at each redshift we plot the mean SFRD over the previous $10^7~{\rm yr}$.
Varying these parameters leads to significant changes in the large-scale evolution of the first stars. We find the most significant changes come from varying $f_{\rm III}$ or $t_{\rm delay}$. Increasing or decreasing the Pop III star formation efficiency leads to more or less Pop III stars, respectively. By $z=20$, there is roughly a two orders of magnitude difference in the Pop III SFRD for $f_{\rm III}=0$ (one Pop III star in each halo) and $f_{\rm III}=0.005$. We note that higher Pop III star formation efficiency also corresponds to a modest decrease in metal-enriched SFR density. This is due to enhanced ionization feedback from extra Pop III photon sources. It has been suggested that there is a self-regulation effect which causes the Pop III SFRD to not depend strongly on the Pop III star formation efficiency \citep[e.g.][]{2000ApJ...534...11H}. This arises because an increase in Pop III star formation efficiency elevates the LW background, which raises the minimum mass of star forming halos and reduces the Pop III SFRD. We do not see this effect in the top left panel of Figure \ref{parameters} because the LW background is caused mainly by metal-enriched stars for most of the redshift range considered in our fiducial model. However, this effect would be seen if we simultaneously increased $t_{\rm delay}$ (discussed below) and $f_{\rm III}$.

Increasing $t_{\rm delay}$ leads to a much longer initial period completely dominated by Pop III star formation. This case corresponds to more disruptive supernovae explosions caused by more massive Pop III stars. In general we find, metal-enriched star formation dominates shortly after the first sites of Pop III star formation recover (after $t_{\rm delay}$).  We note that if metal-enriched star formation is not included in minihalos at all (i.e. only permitted in atomic cooling halos) as was assumed in \cite{2015MNRAS.453.4456V}, the results look qualitatively very similar to increasing $t_{\rm delay}$ as in Figure \ref{parameters}.

In the bottom left panel of Figure \ref{parameters}, we plot the effect of external metal enrichment, which significantly increases the metal-enriched SFRD and modestly decreases the abundance of Pop III stars by polluting pristine halos. We also tested external metal enrichment with $Z_{\rm crit} =  10^{-6}~Z_\odot$, which corresponds to the case with dust cooling from \cite{2005ApJ...626..627O}. This gives essentially identical results as the fiducial value of $Z_{\rm crit} =  10^{-3.5}~Z_\odot$.

The impact of ionization feedback is shown in the bottom right panel of Figure \ref{parameters}. For redshifts greater than $z \sim 25$, ionization feedback can change the metal-enriched SFRD by roughly an order of magnitude, while at lower redshifts it makes closer to a factor of two difference in the metal-enriched SFRD. The increased metal-enriched SFRD due to turning off ionization feedback lowers the amount of Pop III stars initially due to increased LW radiation (see also Figure \ref{tot_popIII}).

\begin{figure}
\includegraphics[width=88mm]{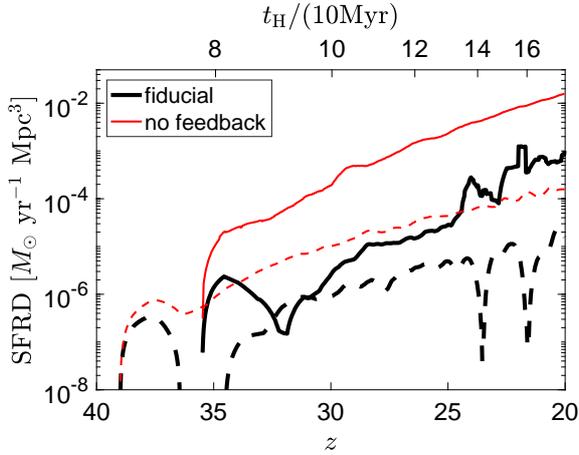}
\caption{\label{no_feedback}
Pop III (dashed curves) and metal-enriched (solid curves) SFRDs in the fiducial case and the case with no LW, external metal enrichment, or ionization feedback. The SFRDs have been smoothed on a scale of $10^7$ yr. } 
\end{figure}

\begin{figure}
\includegraphics[width=88mm]{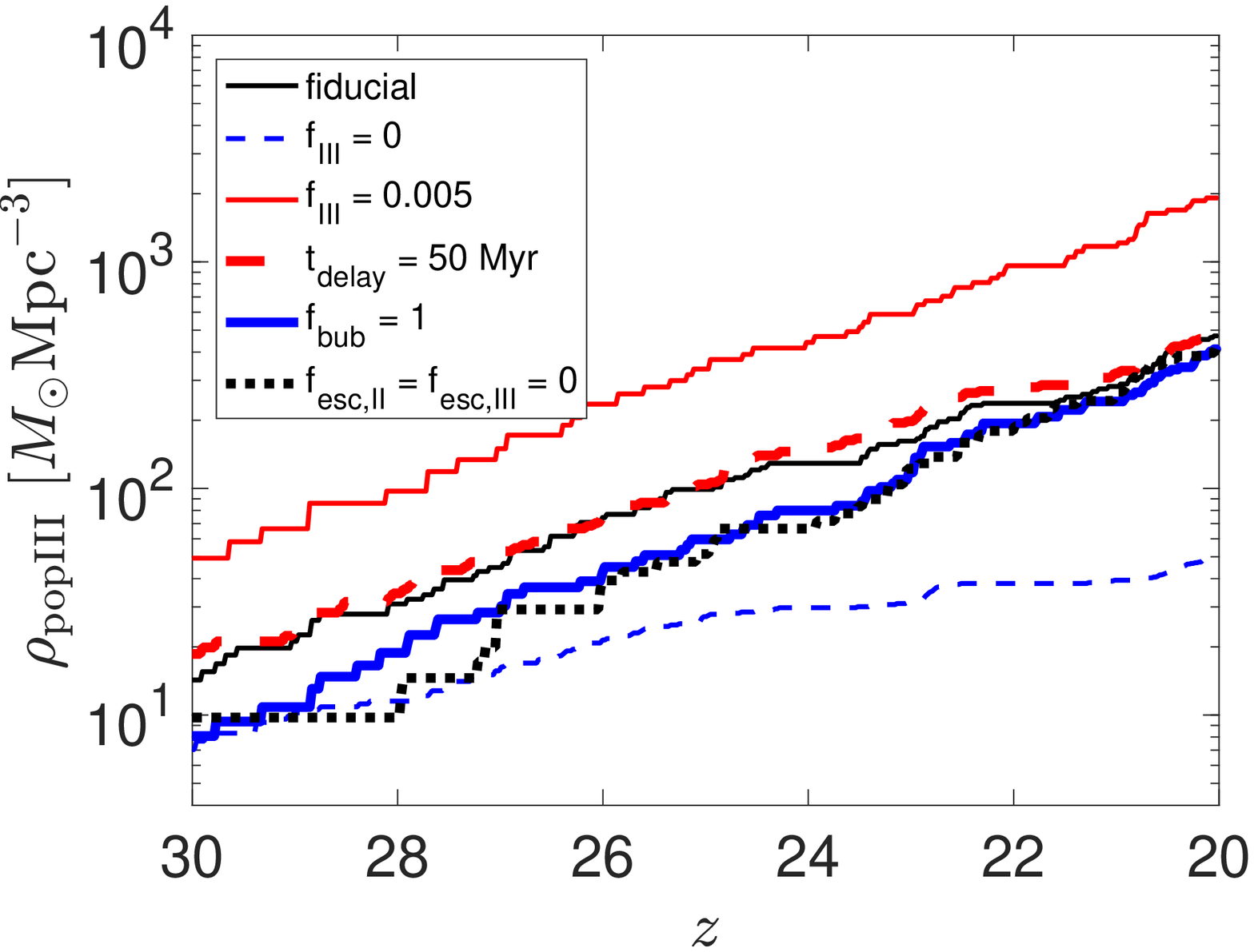}
\caption{\label{tot_popIII}
Cumulative density of Pop III stars formed as a function of cosmic time for various model parameter values. The solid black curve is the fiducial model, and other curves have parameters varied from this model as indicated in the legend. } 
\end{figure}

In Figure \ref{no_feedback}, we show the SFRD density when no feedback is included (i.e. no LW/ionizing feedback or external metal enrichment). In this case, both Pop III and metal-enriched SFRDs are approximately an order of magnitude or more higher than our fiducial model. We note that the difference between the fiducial and the no feedback case can be caused by either LW or ionizing feedback. Including either of these processes returns the Pop III/metal-enriched SFRDs much closer to the fiducial case.  

Note that jagged fluctuations with time in the SFRDs plotted in Figures \ref{parameters}, \ref{no_feedback}, and \ref{clustering} are due to the finite size of our simulation box. For a much larger box, with many more star-forming halos, we would expect smoothed versions of these curves.

We have also computed the total cumulative abundance of Pop III stars, as shown in Figure \ref{tot_popIII}. We note that different parameterizations of our model can change this overall amount by an order of magnitude or more. Unsurprisingly, the Pop III star formation efficiency has the largest impact on the total Pop III abundance. External metal enrichment can reduce the abundance due to polluting potential metal-free sources and increasing the LW feedback by producing more metal-enriched stars. Increasing $t_{\rm delay}$ leads to slightly more Pop III stars initially because there are not metal-enriched stars which increase the LW background, suppressing Pop III star formation.

\begin{table*}
\centering
\caption{\label{table} Physical parameters entering the semi-analytic model and their fiducial values.}
\begin{tabular}{c l| c || c }
Parameter & Description & Fiducial Value  \\
\hline
$f_{\rm III}$ & Pop III star formation efficiency &  0.001 \\[1ex]
$f_{\rm II}$ & Metal-enriched star formation efficiency &  0.05 \\[1ex]
$\eta_{\rm II}$ & LW/Ionizing photons per baryon of metal-enriched stars formed &  4000 \\[1ex]
$\eta_{\rm III}$ & LW/Ionizing photons per baryon of Pop III stars formed &  65000 \\[1ex]
$M_{\rm PopIII}$ & Mass of individual Pop III stars &  $40~M_\odot$ \\[1ex]
$t_{\rm PopIII}$ & Pop III stellar lifetime & $4\times10^6$ yr \\[1ex]
$t_{\rm delay}$ & Delay until metal-enriched star formation after Pop III stars form &  $10^7$ yr \\[1ex]
$t_{\rm PopII}$ & Time metal-enriched stars emit radiation &  $5\times 10^6$ yr \\[1ex]
$Z_{\rm crit} $ & Critical metallicity in externally metal enriched halos  &  $10^{-3.5}~Z_\odot$ \\[1ex]
$M_{\rm min, met} $ & Critical mass in externally metal enriched halos  &  $2\times 10^5~M_\odot$ \\[1ex]
$M_{\rm ion} $ & Ionization feedback mass &  $1.5\times 10^8  \left ( \frac{1+z}{11} \right )^{-3/2}~M_\odot$ \\[1ex]
$f_{ \rm esc, II}$ & Ionizing photon escape fraction in metal-enriched halos &  0.1 \\[1ex]
$f_{ \rm esc, III}$ & Ionizing photon escape fraction in Pop III halos &  0.5 \\[1ex]
$f_{\rm bub}$ & External metal enrichment &  0 (no external enrichment) \\[1ex]
\end{tabular}
\end{table*}

\subsection{Impact of clustering}
Next, we investigate the impact of including/excluding spatial clustering in the feedback prescriptions. For LW feedback, when clustering is excluded, only the global large-scale component, $J_{\rm bg}$, is used. For ionization feedback, ionizing radiation from external halos is no longer considered, however feedback can still occur if a halo (and its progenitors) can ionize a bubble larger than its virial radius, as described above. When clustering is excluded, external metal enrichment is completely disabled. The case without clustering is meant to represent what could be done with merger trees, but without the positions of halos.

\begin{figure*}
\includegraphics[width=88mm]{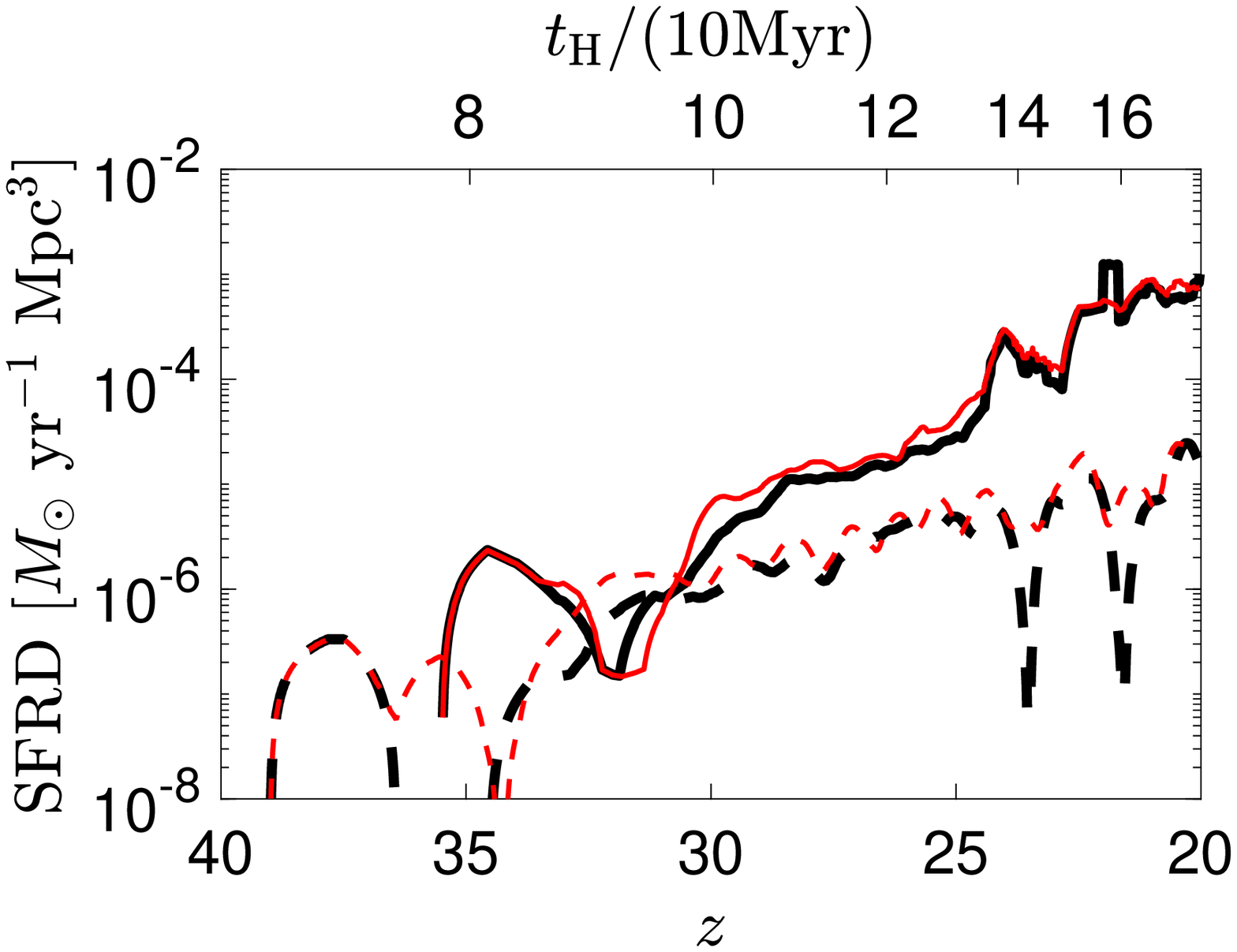}
\includegraphics[width=88mm]{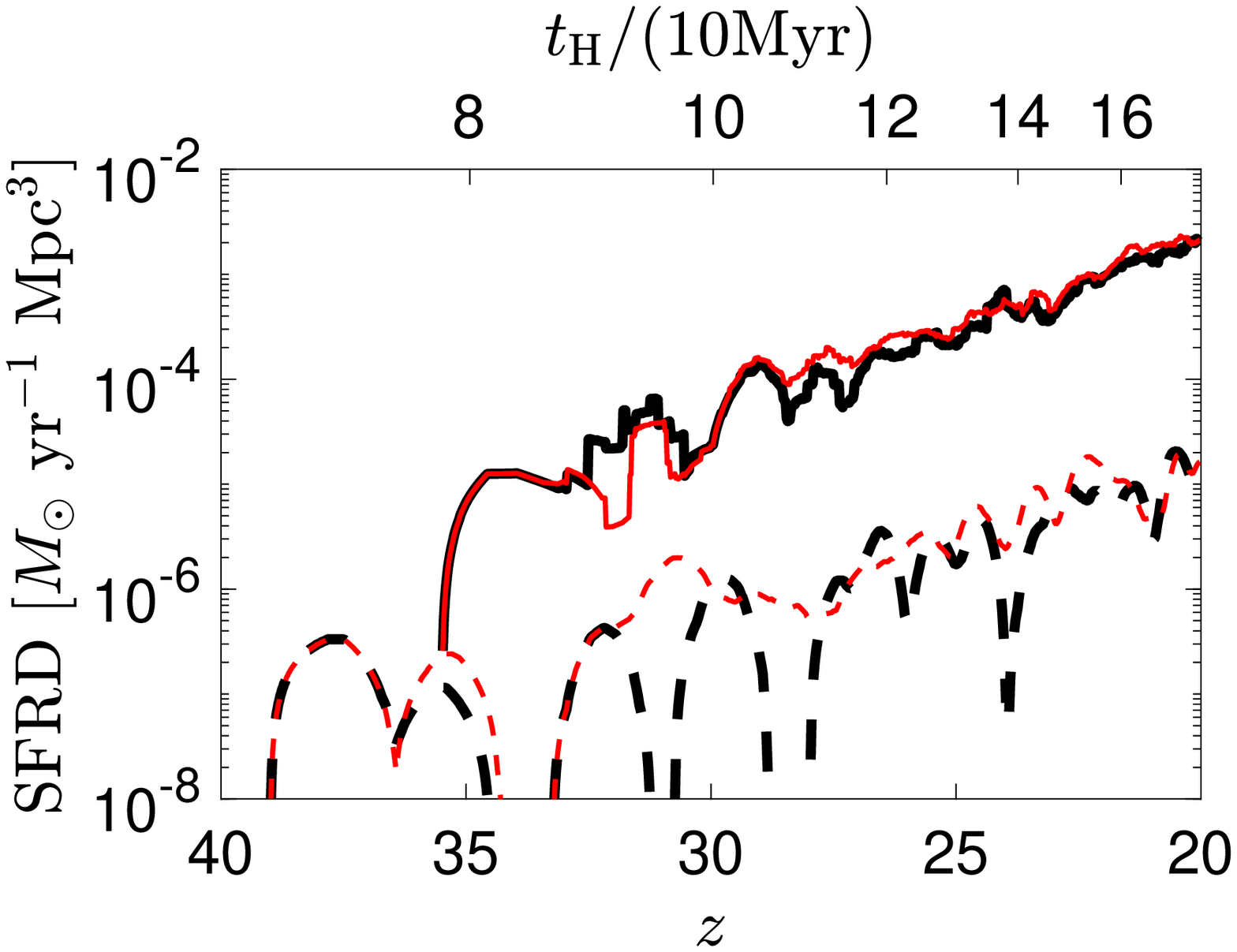}
\caption{\label{clustering}
Pop III (dashed curves) and metal-enriched (solid curves) SFRDs in the cases with (black curves) and without clustering (red curves). The SFRDs have been smoothed on a scale of $10^7$ yr. The case without clustering is meant to illustrate what could be done with merger histories alone (i.e. without running N-body simulations needed to specify halo locations). The left panel is our fiducial model and the right includes LW feedback but no ionization feedback (i.e. $f_{\rm esc, II} = f_{\rm esc,III} = 0$).} 
\end{figure*}

In Figure \ref{clustering}, we show the impact of clustering in the fiducial model and the case with LW feedback, but no ionizing feedback. We find that in both instances the effects of clustering are relatively mild, generally only causing a factor of a two or less difference in the metal-enriched and Pop III SFRDs for the redshifts explored.

The impact of clustering on LW feedback is small because most of the radiation comes from scales larger than our box. While clustering can alter the detailed star formation history of individual halos, it is not important for the global evolution. However, we note though that our simulations do not include clustering on scales larger than the simulation box, as discussed below, which could have some impact. 

For ionization feedback, clustering accounts for a small portion of the feedback. The difference in turning off ionization completely (bottom right panel of Figure \ref{parameters}) and only turning off clustering (left panel of Figure \ref{clustering}) is due to ionizing bubbles created by progenitor halos. Even in the absence of clustering, a halo's progenitors can ionize its location, preventing subsequent star formation. We note that because of this behavior, our ``no clustering'' case does contain information related to positions of halos' progenitors (for the sake of ionization feedback, we assume they were in the same location). Thus, while our no clustering case represents what could be done with merger trees alone (and no halo positions), it implicitly contains more spatial information than most models based on the extended Press-Schechter formalism.

We note that \cite{2006ApJ...649..570K} predict a larger impact of clustering than we find in the present work (though a direct comparison cannot be made since the clustering/no clustering cases are defined differently). This is partly because  \cite{2006ApJ...649..570K} only include ionization feedback and not LW feedback. In our model, we find that the impact of clustering is suppressed by LW feedback, since many of the small minihalos which would be affected by clustered ionizing sources already have their star formation suppressed by the LW background. We verify this by computing a model with ionization feedback and no LW feedback. In this case, we find that the clustering/no clustering models have approximately an order of magnitude difference in the Pop III SFRD below $z\sim 34$. We point out that the total ionized fraction of the IGM is low at the redshifts explored (less than one percent at $z=20$). We expect the impact of clustering on ionization feedback to be larger at lower redshifts.

We note that the effect of external enrichment on the metal-enriched and Pop III SFRDs is due entirely to clustering.

\section{Comparison with Previous Work}
In this section, we discuss our method and results within the context of previous semi-analytical models of the large-scale distribution of Pop III stars. We emphasize that the most novel aspect of our work is that it simultaneously includes inhomogeneous treatments of feedback from LW radiation, ionizing radiation, and external metal enrichment. 

\cite{2009ApJ...700.1672T} use a semi-analytic model based on N-body simulations to estimate the abundance of Pop III stars.
They explored models with inhomogeneous metal enrichment (but constant LW background and no ionizing feedback) and found significant Pop III star formation at late times ($z < 6$).

 \cite{2012MNRAS.425.2854A} also explored the abundance of Pop III stars and so-called ``direct collapse black holes'' using semi-analytic models based on N-body simulations. Their model includes spatial fluctuations in the LW intensity, but does not include spatially inhomogeneous reionization effects or external metal enrichment. We find similar values for the global Pop III SFRD, but a much earlier emergence of dominant metal enriched star formation. This is likely due to their assumption that metal enriched star formation can only occur in dark matter halos more massive than $10^8~M_\odot$.

Another semi-analytic model based on N-body simulations was presented in \cite{2013ApJ...773..108C} to explore the high-redshift abundance of Pop III and metal enriched stars. Their model is similar to ours, but does not include reionization feedback, spatial fluctuations in LW intensity, or external metal enrichment. We find that our fiducial model gives SFRDs (both Pop III and metal enriched) that roughly agree with this work. 

Additionally, there have been a number of works utilizing semi-analytic models to make predictions for Pop III stars which could be found in the Milky Way at $z=0$ \citep[e.g.][]{2016arXiv161100759G, 2016ApJ...826....9I, 2017MNRAS.465..926D, 2017MNRAS.469.1101G, 2017arXiv170607054M}. The comparison to these papers is not straightforward, since the N-body simulations are zoom-in runs on Milky Way-sized halos rather than on unbiased patches of the Universe (i.e.~with matter density equal to the cosmic mean). Additionally, these studies generally focus on lower redshifts than we explore. In future work, we will apply our model to lower redshifts and ultimately to present day Milky Way-like galaxies.

Semi-analytic models built on N-body simulations have been used in the DRAGONS simulation suite to study galaxy formation and reionization in the high-redshift Universe \citep[see e.g.][]{2016MNRAS.459.3025P, 2016MNRAS.462..250M}. However, in order to model the large scales needed for reionization, a lower dark matter resolution is adopted, the smallest minihalos are not resolved, and Pop III star formation is not included. 

We also note that \cite{2017arXiv171002528M} recently developed semi-analytic simulations of Pop III stars using an abundance matching technique to track halos (i.e.~without utilizing N-body simulations). They find Pop III SFRDs generally consistent with our results.

\section{Discussion and Conclusions}
We have developed a semi-analytic model of the formation of the first stars. Our technique utilizes merger trees and three-dimensional spatial information from dark matter only cosmological simulations. We self-consistently populate the dark matter halos from these simulations with Pop III and metal-enriched stars using physically motivated prescriptions for star formation and feedback. 

This approach is intended to provide a middle ground between analytic work, which does not include spatial information or individual halo merger histories, and hydrodynamic cosmological simulations, which include detailed physics but are numerically expensive and have limited dynamic range. Our model includes LW and ionizing feedback, as well as external metal enrichment due to supernovae winds. A number of different physical parameters are included in the model, and can be varied to account for uncertainties in the properties of high-redshift Pop III and metal-enriched star formation.   

We have used our model to study the evolution of the SFRD of the first Pop III and metal-enriched stars in the Universe. We find that varying the model parameters leads to qualitative differences in the global features of the star formation history. The Pop III star formation efficiency, $f_{\rm III}$, and the recovery time required for gas to fall back into a halo and form metal-enriched stars after Pop III supernovae, $t_{\rm delay}$, are especially important. Altering $t_{\rm delay}$ can greatly change the duration of time in which Pop III stars dominate, and unsurprisingly, the relative Pop III and metal-enriched star formation efficiencies determine the importance of Pop III stars in the early Universe. 

Our model was also utilized to study the impact of halo clustering on the evolution of the high-redshift SFRD (i.e. the importance of the three-dimensional positions of individual halos). We find that the effect of small-scale clustering on LW feedback is modest. This is because  most of the flux comes from the background LW component emitted on spatial scales larger than the size of our simulation box. While clustering does expose a small fraction of minihalos to very high LW flux, it does not greatly change the evolution of the SFRD. 

We also find that clustering has a relatively small impact when ionization feedback is included at the redshifts explored ($z \gtrsim 20$). This is because the ionization fraction in the box is quite low at these cosmic times and many halos which would be affected by ionizing feedback already have their star formation suppressed by LW radiation. Additionally, clustering is not very important because this feedback is mainly caused by progenitors of the halos where star formation is suppressed, and thus can be well-approximated without the positions of individual halos. However, as discussed above, ionization feedback may not be accurately captured with analytic work based on the Press-Schechter formalism which does not include merger trees. At lower redshifts, when the total ionization fraction goes up during reionization, we expect the importance of clustering to increase.  

Clustering information is necessary for including external metal enrichment, which we find can increase the metal-enriched SFRD and decrease Pop III star formation at high redshift. 

There are a number of ways our model can be improved in subsequent work. It is relatively straightforward to extend the star formation and feedback prescriptions, making them much more sophisticated. This can include adding Pop III stars of different masses randomly sampled from an assumed IMF. It will also be possible to vary the supernovae recovery time from halo to halo depending on both the masses of the Pop III stars and the masses of the minihalos. Additionally, it would be advantageous to introduce a more detailed prescription for metal-enriched star formation. For example, it may be important to include changes in star formation due to mergers of star-forming halos or additional negative feedback on star formation due to supernovae explosions. We note however that even though we use a very simple prescription for star formation, our model agrees reasonably well with more sophisticated simulations. In our fiducial model, we find that typical metal enriched halos at $z=20$ have halo masses of $M_{\rm vir}\sim 10^7~M_\odot$ and total metal-enriched stellar masses of $M_*\sim3000~M_\odot$, with a factor of a few scatter in each of these quantities, in general agreement with \cite{2015ApJ...807L..12O}.

We note that one shortcoming of our technique is that the N-body box size is much smaller than the horizon of LW photons contributing to the LW background (which comes from scales of $\approx 100~{\rm Mpc}$). Thus, while our calculation of the LW background is self-consistent, it does not take into account density fluctuations on scales larger than the box. In future work, it may be possible to combine the model described here with large-scale semi-numerical models such as \cite{2012Natur.487...70V}, which do not track individual halos, but follow the total SFR in voxels a few comoving Mpc in size. 

We also note that the baryon-dark matter streaming velocity \citep{2010PhRvD..82h3520T, 2012MNRAS.424.1335F}, which can suppress star formation in minihalos, has not been included in the present work. Thus, our results represent a region with low streaming velocity, however the streaming velocity would reduce the background component of the LW flux computed here. It may be possible to address this by combining a number of different semi-analytic simulations that have different values of the local streaming velocity, but that is beyond the scope of this paper. 

We have also assumed a LW escape fraction of unity throughout this paper. We do not expect this assumption to have a large impact on our results, however for the case of $f_{\rm III}=0$, we may overestimate the amount of LW emission which escapes the halos \citep{2015MNRAS.454.2441S}.
 
In future work, we plan to compare our model directly with hydrodynamical cosmological simulations. Matching to detailed simulations will help to identify the most important physical processes which determine the global properties of first stars. We note that, even without any tuning, our fiducial model is generally consistent with more detailed hydrodynamical cosmological simulations \citep[e.g.~see figure 2 in][]{2013ApJ...773...83X}. We expect a closer match to be achievable by adjusting the parameters and modestly improving of our model (i.e.~by adding a more realistic Pop III IMF and halo mass/redshift dependent metal-enriched star formation efficiency).

Given the relatively low numerical cost of this semi-analytic method, it will be valuable for exploring a large range of the highly uncertain parameter space related to the first stars. Potential applications include predictions for high-redshift cosmological 21cm observations, mergers of Pop III remnants detectable with gravitational waves, stellar archaeology in the milky way, and an early partial reionization which can be inferred from observations of the CMB.

\section*{Acknowledgements}
The Flatiron Institute is supported by the Simons Foundation. 
ZH was supported by NASA grant NNX15AB19G and by a
Simons Fellowship in Theoretical Physics.  GLB was supported by NASA
grant NNX15AB20G and NSF grants AST-1312888 and AST-1615955. The required N-body simulations were run on the NASA Pleiades supercomputer.

\bibliography{paper.bib}
\end{document}